\documentclass[english]{IEEEtran}
\usepackage[T1]{fontenc}
\usepackage{float}
\usepackage{amsmath}
\usepackage{amssymb}
\usepackage{graphicx}
\usepackage{xcolor}
\usepackage{amsthm}
\usepackage{pdfcolmk}
\usepackage{amsthm}
\usepackage{setspace}
\usepackage{wasysym}
\PassOptionsToPackage{normalem}{ulem}
\usepackage{ulem}
\makeatletter

\floatstyle{ruled}
\newfloat{algorithm}{tbp}{loa}
\providecommand{\algorithmname}{Algorithm}
\floatname{algorithm}{\protect\algorithmname}

\makeatother

\usepackage{babel}
\begin{document}
\title{Two-Timescale Joint Transmit and Pinching Beamforming for Pinching-Antenna
Systems}
\author{{\normalsize Luyuan Zhang, Xidong Mu, }{\normalsize\textit{Member,
IEEE}}{\normalsize , An Liu, }{\normalsize\textit{Senior Member, IEEE
}}{\normalsize and Yuanwei Liu, }{\normalsize\textit{Fellow IEEE}}{\normalsize\thanks{L. Zhang and A. Liu are with the College of Information Science and
Electronic Engineering, Zhejiang University, Hangzhou 310027, China
(email: \{22131107, anliu\}@zju.edu.cn).}\thanks{X. Mu is with the Centre for Wireless Innovation (CWI), Queen's University
Belfast, Belfast, BT3 9DT, U.K. (x.mu@qub.ac.uk).}\thanks{Y. Liu is with the Department of Electrical and Electronic Engineering
(EEE), The University of Hong Kong, Hong Kong. (yuanwei@hku.hk).}}}
\maketitle
\begin{abstract}
Pinching antenna systems (PASS) have been proposed as a revolutionary
flexible antenna technology which facilitates line-of-sight links
via numerous low-cost pinching antennas with adjustable activation
positions over waveguides. This letter proposes a two-timescale joint
transmit and pinching beamforming design for the maximization of sum
rate of a PASS-based downlink multi-user multiple input single output
system. A primal dual decomposition method is developed to decouple
the two-timescale problem into two sub-problems: 1) A Karush-Kuhn-Tucker-guided
dual learning-based approach is proposed to solve the short-term transmit
beamforming design sub-problem; 2) The long-term pinching beamforming
design sub-problem is tackled by adopting a stochastic successive
convex approximation method. Simulation results demonstrate that the
proposed two-timescale algorithm achieves a significant performance
gain compared to other baselines.
\end{abstract}


\section{Introduction\label{sec:introduction}}

Flexible-antenna techniques such as reconfigurable intelligent surfaces
(RISs) \cite{mu2021simultaneously}, movable antennas \cite{zhu2023modeling},
and fluid antennas \cite{new2024tutorial}, have been developed to
break the limitation of fixed-channel assumptions in the sixth generation
(6G) and beyond wireless network. Nevertheless, existing flexible-antenna
techniques have several limitations, e.g., RIS faces double fading
issue \cite{ozdogan2019intelligent} and the movable/fluid antenna
has limited range of antenna movement. Moreover, existing flexible-antenna
techniques are usually equipped with a fixed number of antennas, which
leads to limited reconfigurability. 

To tackle these drawbacks, pinching antenna systems (PASS) have recently
emerged as a revolutionary flexible-antenna technique \cite{suzuki2022pinching},
which is built on a rod-like transmission line, named dielectric waveguide.
The dielectric waveguide acts as a leaky-wave antenna, where multiple
radiation points, implemented by dielectric paricals, known as pinching
antennas (PAs), can be flexibly pinched and released anywhere along
this waveguide. This feature provides a low-cost and scalable approach
to implementing multiple input multiple output (MIMO) while also facilitating
so-called \textit{pinching beamforming}. Among existing flexible antenna
techniques, PASS has several unique advantages, e.g., PASS can greatly
reduce the large-scale path loss to deliver wireless services over
meter-order distances, and maintain stable LoS links to avoid blockage
especially for high-frequency communications. 

Motivated by the above benefits, some researchers have begun to explore
PASS-enabled wireless communications. For example, the authors of
\cite{ding2024flexible} studied both single-waveguide and multiple-waveguide
PASS, and also analyzed the performance upper bound. The authors of
\cite{tegos2025minimum} focused on maximizing the minimum achievable
data rate in the uplink PASS, and separately optimizes the positions
of the PAs by transferring the problem into a convex one. The authors
of \cite{wang2025modeling} proposed a penalty-based alternating optimization
algorithm for the joint optimization of transmit and pinching beamforming
under both continuous and discrete pinching antenna activations. In
\cite{xu2025joint}, the authors proposed both optimization-based
and learning-based methods to solve the joint optimization problem. 

However, aforementioned works focus on the single-timescale optimization,
where the pinching beamforming and/or the transmit beamforming are
optimized simultaneously in a real-time manner, which is hard to achieve
due to hardware limitations. In practice, it is challenging to adjust
the position of PAs in real-time. To circumvent this drawback, in
this letter, we propose a two-timescale joint transmit and pinching
beamforming design for the maximization of sum rate of a PASS-based
downlink multi-user multiple input single output (MISO) communication
system. In this framework, the pinching beamforming is viewed as a
long-term variable, which is optimized at the beginning of each channel
statistics coherence time block, while the transmit beamforming is
optimized based on a fixed pinching beamforming and the short-term
instant channel state information (CSI). For the joint beamforming
design, we develop a primal dual decomposition (PDD) method to decouple
the problem into the short-term transmit beamforming design sub-problem
and long-term pinching beamforming design sub-problem. Then, a Karush-Kuhn-Tucker
(KKT)-guided dual learning (KDL)-based approach is proposed to obtain
a stationary point of the short-term transmit beamforming and the
corresponding gradient for the long-term variables is attained by
back propagation (BP). The long-term stochastic pinching beamforming
optimization is solved by a stochastic successive convex approximation
(SSCA) method based on the values and gradients obtained using the
short-term samples. Simulation results demonstrate that the proposed
two-timescale algorithm achieves a significant performance gain compared
to other baselines.

\section{System Model And Problem Formulation \label{sec:System-model}}

\begin{figure}[tp]
\begin{centering}
\textsf{\includegraphics[scale=0.55]{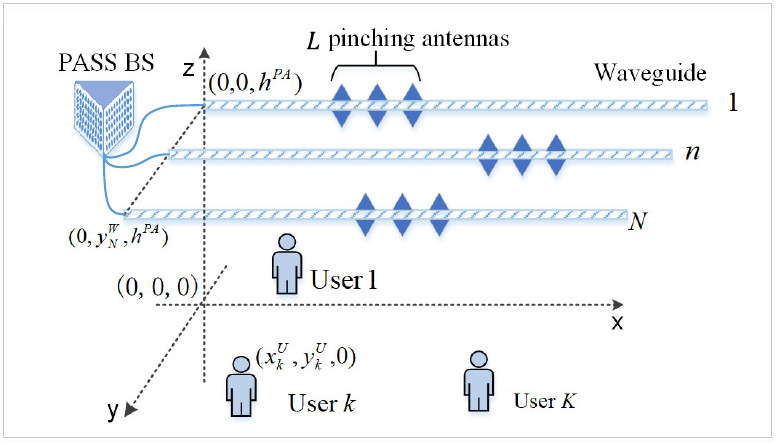}}
\par\end{centering}
\caption{\label{fig:system-model}An illustration for the considered PASS-based
downlink system.}
\end{figure}

As shown in Fig. \ref{fig:system-model}, we consider a downlink PASS-based
downlink MISO communication system. The base station (BS) serves a
set of $\mathcal{K}=\{1,...,K\}$ single-antenna users with $N$ waveguides.
$L$ PAs are mounted on each waveguide to jointly serve the users,
and therefore there are $M=L\times N$ PAs in total. To ensure multiplexing
gains, we assume $N=K$. It is assumed that all waveguides are installed
at a fixed height of $h^{\textrm{PA}}$, and the resulting PASS spans
across a rectangular area with a size of $S_{x}\times S_{y}$ $\textrm{m}^{2}$.
The three-dimension Cartesian coordinate of the feed point for the
$n$-th waveguide is given by $\boldsymbol{\eta}_{n}^{\textrm{W}}=\left[0,y_{n}^{\textrm{W}},h^{\textrm{PA}}\right]^{T}$,
where $y_{n}^{\textrm{W}}$ is the $y$-axis coordinate of this waveguide.
The location of each PA $a_{n,l}$ along with the $n$-th waveguide
can be defined as $\boldsymbol{\eta}_{n,l}\left(x_{n,l}\right)=\left[x_{n,l},y_{n,l},h^{\textrm{PA}}\right]^{T}$,
where $x_{n,l}$ is the adjustable pinched location of PA $a_{n,l}$
along $x$-axis, and $y_{n,l}$ is the fixed coordinate along $y$-axis
which depends on the corresponding waveguide deployment, i.e., $y_{n,l}=y_{n}^{\textrm{W}}$.
Let $\mathbf{x}_{n}=\left[x_{n,1},...,x_{n,L}\right]^{T}\in\mathbb{R}^{L\times1}$
denote the $x$-axis locations of PAs over the $n$-th waveguide,
which satisfy $0\leq x_{n,1}<...<x_{n,L}\leq S_{x}$, $\forall n\in\mathcal{N}.$
$\mathbf{X}=\left[\mathbf{x}_{1};...;\mathbf{x}_{N}\right]\in\mathbb{R}^{N\times L}$
denotes the locations of all PAs. Moreover, the location of user $k$
is denoted as $\boldsymbol{\eta}_{k}^{\textrm{U}}=\left[x_{k}^{\textrm{U}},y_{k}^{\textrm{U}},0\right]^{T},\forall k\in\mathcal{K}$.

\subsection{Signal Model}

By activating PAs at different points along the waveguides, both the
phases of incident signals and the large-scale fading can be altered.
Firstly, we denote $\mathbf{G}\left(\mathbf{X}\right)\in\mathbb{C}^{M\times N}$
as the path response when the signals are propagated from the feed
point of each waveguide to specific PAs. $\mathbf{G}\left(\mathbf{X}\right)$
is a block-diagonal matrix, which can be given by 
\begin{equation}
\mathbf{G}\left(\mathbf{X}\right)=\left[\begin{array}{cccc}
\mathbf{g}_{1}(\mathbf{x}_{1}) & \mathbf{0} & ... & \mathbf{0}\\
\mathbf{0} & \mathbf{g}_{2}(\mathbf{x}_{2}) & ... & \mathbf{0}\\
... & ... & ... & ...\\
\mathbf{0} & \mathbf{0} & ... & \mathbf{g}_{N}(\mathbf{x}_{N})
\end{array}\right],
\end{equation}
where $\mathbf{g}_{n}(\mathbf{x}_{n})\in\mathbb{C}^{L\times1}$ denotes
the response vector from the feed point of waveguide $n$ to the associated
subset of PAs. The $l$-th element of $\mathbf{g}_{n}(\mathbf{x}_{n})$
can be expressed as 
\begin{equation}
g_{n,l}(x_{n,l})=\frac{1}{\sqrt{L}}e^{-j2\pi\frac{||\boldsymbol{\eta}_{n,l}\left(x_{n,l}\right)-\boldsymbol{\eta}_{n}^{\textrm{W}}||}{\lambda_{\textrm{W}}}},
\end{equation}
where $\lambda_{\textrm{W}}=\frac{\lambda_{f}}{n_{\textrm{eff}}}$
is the guided wavelength, with $n_{\textrm{eff}}$ denoting the effective
refractive index of the dielectric waveguide \cite{pozar2021microwave}.
Moreover, $||\boldsymbol{\eta}_{n,l}\left(x_{n,l}\right)-\boldsymbol{\eta}_{n}^{\textrm{W}}||=x_{n,l}$
denotes the distance from waveguide $n$ to PA $a_{n,l}$.

User $k$'s channel vector can be expressed as $\mathbf{h}_{k}^{H}\left(\mathbf{X}\right)=\left[\mathbf{h}_{k,1}^{h}(\mathbf{x}_{1}),...,\mathbf{h}_{k,n}^{H}(\mathbf{x}_{N})\right]\in\mathbb{C}^{1\times M}$,
where $\mathbf{h}_{k,n}^{H}(\mathbf{x}_{n})$ denotes the channel
from the $n$-th waveguide to user $k$, which is determined by the
pinching deployment $\mathbf{x}_{n}$. The $l$-th element of $\mathbf{h}_{k,n}^{H}(\mathbf{x}_{n})$
indicates the channel from PA $a_{n,l}$ to user $k$, which is given
by
\begin{equation}
h_{k,n,l}^{H}(x_{n,l})=\frac{\sqrt{\beta}e^{-j\kappa r(x_{n,l},\boldsymbol{\eta}_{k}^{\textrm{U}})}}{r(x_{n,l},\boldsymbol{\eta}_{k}^{\textrm{U}})},
\end{equation}
where $\kappa=2\pi/\lambda_{f}$ denotes the wave-domain number, and
$\lambda_{f}$ is the wavelength. $\beta=\frac{c}{4\pi f_{c}}$ is
a constant with $c$, $f_{c}$ and $\lambda$ denoting the speed of
light, the carrier frequency, and the wavelength in free space, respectively.
$r(x_{n,l},\boldsymbol{\eta}_{k}^{\textrm{U}})=||\boldsymbol{\eta}_{k}^{\textrm{U}}-\boldsymbol{\eta}_{n,l}\left(x_{n,l}\right)||$
denotes the distance between PA $a_{n,l}$ and user $k$.

The signal is multiplexed at the baseband through the conventional
transmit beamforming $\mathbf{W}$, and then converted by the RF chain
and fed into the waveguide for radiation. Therefore, the transmitted
signal of user $k$ radiated by the PASS to the free space can be
expressed as 
\begin{equation}
\mathbf{s}_{k}=\mathbf{G}\left(\mathbf{X}\right)\mathbf{w}_{k}c_{k}\in\mathbb{C}^{M\times1},
\end{equation}
where $\mathbf{w}_{k}$ is the conventional transmit beamforming vector
for user $k$ and $c_{k}$ denotes the normalized data symbol, i.e.,
$\mathbb{E}\left[|c_{k}|^{2}\right]=1$. Hence, the received signal
at user $k$ through the channels manipulated by PAs is expressed
as:
\begin{equation}
y_{k}=\mathbf{h}_{k}^{H}\left(\mathbf{X}\right)\mathbf{G}\left(\mathbf{X}\right)\sum_{k=1}^{K}\mathbf{w}_{k}c_{k}+n_{k},
\end{equation}
where $n_{k}\sim\mathcal{CN}\left(0,\delta^{2}\right)$ indicates
the additive white Gaussian noise (AWGN) at user $k$ with zero mean
and variance $\delta^{2}$. 

The signal-to-interference-plus-noise ratio (SINR) $\textrm{SINR}_{k}$
for user $k$ is be given by
\begin{align}
\textrm{SINR}_{k} & =\frac{|\mathbf{h}_{k}^{H}\mathbf{G}\mathbf{w}_{k}|^{2}}{\sum_{i\neq k}|\mathbf{h}_{k}^{H}\mathbf{G}\mathbf{w}_{i}|^{2}+\sigma^{2}},\forall k\in\mathcal{K}.
\end{align}

As a result, the achievable data rate $R_{k}$ for user $k$ can be
expressed as $R_{k}=\log_{2}\left(1+\textrm{SINR}_{k}\right)$,$\forall k\in\mathcal{K}.$ 

\subsection{Two-Timescale Sum Rate Maximization Problem Formulation}

\begin{figure}[tp]
\begin{centering}
\textsf{\includegraphics[scale=0.55]{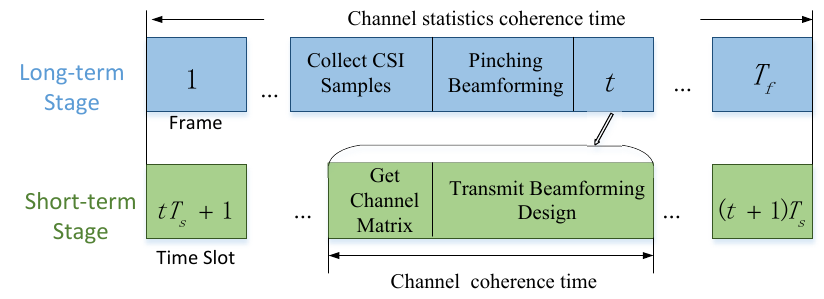}}
\par\end{centering}
\caption{\label{fig:Two-stage}Two-timescale scheme for joint transmit and
pinching beamforming design.}
\end{figure}

In practice, it is challenging to adjust the position of pinching
antennas in real-time due to the restriction of hardware. Therefore,
we introduce a two-timescale transmission scheme, depicted in Fig.
\ref{fig:Two-stage}. Specifically, a channel statistics coherence
time block is considered during which channel statics are assumed
to remain nearly constant. The optimization process is divided into
two timescales, where the pinching beamforming and transmit beamforming
are optimized on different timescales. In the long-term stage, the
BS collects a small amount of instant CSI samples $\left\{ \mathcal{H}\left(n_{s}\right)\right\} _{n_{s}=\left\{ 1,...,N_{s}\right\} }$,
where $N_{s}$ is the number of samples and $\mathcal{H}\left(n_{s}\right)\triangleq\left\{ \mathbf{h}_{k}\right\} _{k\in\mathcal{K}}$
represents the $n_{s}$-th sample which contains the set of user's
CSI. Then, the pinching beamforming is designed by optimizing $\mathbf{X}$
based on these samples. In the short-term stage, the long-term locations
of  pinching antennas remain unchanged, while for each time slot,
the BS acquires the channel matrix and applies the short-term algorithm
to optimize the conventional beamforming vectors $\mathbf{W}=\left[\mathbf{w}_{k}\right]_{k\in\mathcal{K}}$. 

The goal of this paper is to maximize the average downlink sum rate
of users across a channel statistics coherence time block, by jointly
optimizing the long-term pinching beamforming $\mathbf{X}$, and the
conventional beamforming vectors $\mathbf{W}(t)$. The associated
optimization problem can be formulated as 
\begin{subequations}
\begin{equation}
\mathcal{P}_{0}:\max\,_{\mathbf{X}}\mathbb{E}_{\mathcal{H}}\left\{ \max_{\mathbf{W}}\sum_{k\in\mathcal{K}}R_{k}\right\} \label{eq:two-stage problem}
\end{equation}
\begin{equation}
\textrm{s.t.\ }\sum_{k}||\boldsymbol{w}_{k}||_{2}^{2}\leq P_{\textrm{max}},\label{eq:c1}
\end{equation}
\begin{equation}
|x_{n,l}-x_{n,l-1}|\geq\triangle,\forall n\in\mathcal{\mathcal{N}},\forall1<l\leq L\in\mathcal{\mathcal{N}},\label{eq:c2}
\end{equation}
\begin{equation}
0\leq x_{n,l}\leq S_{x},\forall l\in\mathcal{\mathcal{M}},\label{eq:c5-1}
\end{equation}
\end{subequations}
where Constraint (\ref{eq:c1}) is the total transmission power constraint.
Constraint (\ref{eq:c2}) ensures that the distance between adjacent
pinching antennas is greater than or equal to $\triangle$ to avoid
coupling effects and constraint (\ref{eq:c5-1}) guarantees the location
of each PA does not exceed the maximum length of the connected waveguide.

However, it is challenging to solve $\mathcal{P}_{0}$ due to the
non-convexity of the objective. Moreover, the short-term and long-term
variables are highly coupled with each other in the rate terms. To
address this issue, we will develop a two-timescale joint transmit
and pinching beamforming design in the following.

\section{Two-Timescale Joint Beamforming Design \label{sec:Proposed-Algorithm}}

In this section, we propose a primal-dual decomposition based joint
transmit and pinching beamforming algorithm for the considered downlink
systems. Firstly, the problem (\ref{eq:two-stage problem}) is decomposed
into two sub-problems by adopting the two-stage PDD method in \cite{liu2021two}.
Then, the long-term stage problem is solved by developing SSCA in
\cite{liu2019stochastic}. Finally, we propose a KDL-based learning
approach to solve the short-term stage problem.

\subsection{Primal-Dual Decomposition}

We first decompose $\mathcal{P}_{0}$ into two sub-problems using
PDD method. Specifically, for a fixed long-term variable $\mathbf{X}$,
let $\mathbf{W}^{s}(\mathbf{X};\mathcal{H})$ denote a stationary
point (up to certain tolerable error) of the \textit{short-term }\textit{sub-problem}
$\mathcal{P}_{S}(\mathbf{X};\mathcal{H})$, which is given by
\begin{subequations}
\begin{equation}
\mathcal{P}_{S}:\max_{\mathbf{W}}\sum_{k\in\mathcal{K}}R_{k}(\mathbf{X},\mathbf{W};\mathcal{H})\label{eq:two-stage problem-1}
\end{equation}
\begin{equation}
\textrm{s.t.\ }\sum_{k}||\boldsymbol{w}_{k}||_{2}^{2}\leq P_{\textrm{max}},\label{eq:c1-2}
\end{equation}
\end{subequations}

With the notation of $\mathbf{W}^{s}(\mathbf{X};\mathcal{H})$, we
can formulate the following \textit{long-term sub-problem}:
\begin{subequations}
\begin{equation}
\mathcal{P}_{L}:\min_{\mathbf{X}}f(\mathbf{X})\triangleq\mathbb{E}_{\mathcal{H}}\left\{ -\sum_{k\in\mathcal{K}}R_{k}(\mathbf{X},\mathbf{W}^{s}(\mathbf{X};\mathcal{H});\mathcal{H})\right\} \label{eq:two-stage problem-2}
\end{equation}
\begin{equation}
\textrm{s.t.\ }|x_{n,l}-x_{n,l-1}|\geq\triangle,\forall n\in\mathcal{\mathcal{N}},\forall1<l\leq L\in\mathcal{\mathcal{N}},\label{eq:c5-2}
\end{equation}
\begin{equation}
0\leq x_{n,l}\leq S_{x},\forall l\in\mathcal{\mathcal{M}}.\label{eq:c5-1-1}
\end{equation}
\end{subequations}

According to Theorem 2 in \cite{liu2021two}, a KKT solution can be
found by first solving a stationary point $\mathbf{X}^{*}$ of the
long-term sub-problem, and then finding a stationary point $\mathbf{W}^{s}(\mathbf{X};\mathcal{H})$
of $\mathcal{P}_{S}(\mathbf{X}^{*};\mathcal{H})$ for each sample
$\left\{ \mathcal{H}\left(n_{s}\right)\right\} _{n_{s}=\left\{ 1,...,N_{s}\right\} }$.
\begin{algorithm}
\caption{\label{long-term sub}Long-Term Pinching Beamforming}

\textbf{Initialization:} Initialize $\mathbf{X}^{0}$; $t=0$.

\textbf{Input: $\{\gamma^{t},\rho^{t}\}.$}

\textbf{While} $t\leq T_{f}$:

\phantom{} \phantom{}{\footnotesize 1}. Obtain a mini-batch $\left\{ \mathcal{H}\left(n_{s}\right)\right\} _{n_{s}=\left\{ 1,...,N_{s}\right\} }$;

\phantom{} \phantom{} \phantom{} Construct the surrogate function
$\bar{f}^{t}(\mathbf{X})$ according to (\ref{eq:surrogate function}).

\phantom{} \phantom{}{\footnotesize 2}. Obtain $\bar{\mathbf{X}}^{t}$
by solving (\ref{eq:surrogate-x});

\phantom{} \phantom{}{\footnotesize 3}. Update $\mathbf{X}^{t+1}$
according to (\ref{eq:updatex}).

\phantom{} \phantom{}{\footnotesize 4}.\textbf{ $t=t+1$.}

\textbf{End}
\end{algorithm}

\subsection{Long-Term Pinching Beamforming}

The long-term problem $\mathcal{P}_{L}$ is a single-stage stochastic
optimization problem with non-convex objective function. We adopt
the SSCA framework \cite{liu2019stochastic} to deal with the stochasticity
and non-convexity of $\mathcal{P}_{L}$.

In the $t$-th iteration, as illustrated in Fig. \ref{fig:Two-stage},
one random mini-batch $\left\{ \mathcal{H}\left(n_{s}\right)\right\} _{n_{s}=\left\{ 1,...,N_{s}\right\} }$
of $N_{s}$ channel state samples are first obtained and the surrogate
function $\bar{f}^{t}(\mathbf{X})$ is constructed based on the mini-batch
$\left\{ \mathcal{H}\left(n_{s}\right)\right\} $ and the current
iterate $\mathbf{X}^{t}$. The surrogate function $\bar{f}^{t}(\mathbf{X})$
can be viewed as a convex approximation of the objective function
of the long-term problem $\mathcal{P}_{L}$.

Specifically, a structured surrogate function $\bar{f}^{t}(\mathbf{X})$
can be constructed as \cite{liu2019stochastic}
\begin{align}
\bar{f}^{t}(\mathbf{X}) & =(1-\rho^{t})f^{t-1}+\rho^{t}\frac{1}{N_{s}}\sum_{j=1}^{N_{s}}[g(\mathbf{X}^{t},\mathbf{W}_{j}^{t},n_{s}^{t}(j))\nonumber \\
+ & \partial_{\mathbf{X}}^{T}g(\mathbf{X}^{t},\mathbf{W}_{j}^{t},n_{s}^{t}(j))(\mathbf{X}-\mathbf{X}^{t})]\nonumber \\
+ & ((1-\rho^{t})\mathbf{f}_{x}^{t-1}+\mathbf{f}_{W}^{t})^{T}(\mathbf{X}-\mathbf{X}^{t})\nonumber \\
+ & \tau\left(||\mathbf{X}-\mathbf{X}^{t}||^{2}\right),\label{eq:surrogate function}
\end{align}
where $\mathbf{W}_{j}^{t},$ is an abbreviation for $\mathbf{W}^{s}(\mathbf{X}^{t},n_{s}^{t}(j))$,
$g(\mathbf{X}^{t},\mathbf{W}_{j}^{t},n_{s}^{t}(j))$$=$$-\sum_{k\in\mathcal{K}}R_{k}(\mathbf{X}^{t},\mathbf{W}_{j}^{t},n_{s}^{t}(j))$,
$\{\rho^{t}\in(0,1]\}$ is a decreasing sequence satisfying $\rho^{t}\rightarrow0$,
$\sum_{t}\rho^{t}=\infty$, $\sum_{t}(\rho^{t})^{2}<\infty$, $\tau>0$
can be any constant. $f^{t}$ is an approximation for $f(\mathbf{X}^{t})$
and it is updated recursively according to
\begin{equation}
f^{t}=(1-\rho^{t})f^{t-1}+\rho^{t}\frac{1}{N_{s}}\sum_{j=1}^{N_{s}}g(\mathbf{X}^{t},\mathbf{W}_{j}^{t},n_{s}^{t}(j)),
\end{equation}
with $f^{-1}=0$. $\mathbf{f}_{x}^{t}$ and $\mathbf{f}_{W}^{t}$
are approximations for the gradients $\mathbb{E}_{\mathcal{H}}[\partial_{\mathbf{X}}g(\mathbf{X},\mathbf{W}^{s}(\mathbf{X};\mathcal{H});\mathcal{H})]$
and $\mathbb{E}_{\mathcal{H}}[\partial_{\mathbf{X}}\mathbf{W}^{s}(\mathbf{X};\mathcal{H})\partial_{\mathbf{W}}g(\mathbf{X},\mathbf{W}^{s}(\mathbf{X};\mathcal{H});\mathcal{H})]$,
respectively, which are recursively according to $\partial_{\mathbf{X}}\mathbf{W}\partial_{\mathbf{W}}g$
\[
\mathbf{f}_{x}^{t}=(1-\rho^{t})\mathbf{f}_{x}^{t-1}+\rho^{t}\frac{1}{N_{s}}\sum_{j=1}^{N_{s}}\partial_{\mathbf{X}}g(\mathbf{X}^{t},\mathbf{W}_{j}^{t},n_{s}^{t}(j)),
\]
\begin{equation}
\mathbf{f}_{W}^{t}=(1-\rho^{t})\mathbf{f}_{W}^{t-1}+\rho^{t}\frac{1}{N_{s}}\sum_{j=1}^{N_{s}}\partial_{\mathbf{X}}\mathbf{W}_{j}^{t}\partial_{\mathbf{W}}g(\mathbf{X}^{t},\mathbf{W}_{j}^{t},n_{s}^{t}(j)),
\end{equation}
with $\mathbf{f}_{x}^{-1}=\mathbf{f}_{W}^{-1}=0$, $\partial_{\mathbf{X}}\mathbf{W}_{j}^{t}=\partial_{\mathbf{X}}\mathbf{W}^{s}(\mathbf{X}^{t},n_{s}^{t}(j))$
is the derivative of the vector function $\mathbf{W}^{s}(\mathbf{X}^{t},n_{s}^{t}(j))$
to the vector $\mathbf{X}$.

Then, the optimal solution $\bar{\mathbf{X}}^{t}$ of the following
problem is solved:
\[
\bar{\mathbf{X}}^{t}=\arg\min_{\mathbf{X}}\bar{f}^{t}(\mathbf{X})
\]
\begin{equation}
\textrm{s.t.}(\ref{eq:c5-2}),(\ref{eq:c5-1-1}),\label{eq:surrogate-x}
\end{equation}
which is a convex approximation of $\mathcal{P}_{L}$. Finally, given
$\bar{\mathbf{X}}^{t}$, $\mathbf{X}$ is updated according to 
\begin{equation}
\mathbf{X}^{t+1}=(1-\gamma^{t})\mathbf{X}^{t}+\gamma^{t}\bar{\mathbf{X}}^{t},\label{eq:updatex}
\end{equation}
where $\{\gamma^{t}\in(0,1]\}$ is a decreasing sequence satisfying
$\gamma^{t}\rightarrow0$, $\sum_{t}\gamma^{t}=\infty$, $\sum_{t}(\gamma^{t})^{2}<\infty$,
$\lim_{t\rightarrow\infty}\gamma^{t}/\rho^{t}=0$.

Based on the above, we can solve the long-term sub-problem based on
the SSCA, and the procedure is summarized in Algorithm \ref{long-term sub}.

\subsection{Short-Term Transmit Beamforming \label{subsec:Short-Term-Transmit-Beamforming}}

To avoid confusion with the iteration of the long-term sub-algorithm,
an iteration of the short-term sub-algorithm will be called an \textit{inner
iteration}. The short-term stage sub-problem $\mathcal{P}_{S}$ can
be solved by introducing the mean-square estimation error (MSE) $E_{k}$:
\begin{equation}
E_{k}\triangleq|u_{k}\boldsymbol{h}_{k}^{H}\mathbf{G}\boldsymbol{w}_{k}-1|^{2}+\sum_{i\neq k}|u_{k}\boldsymbol{h}_{k}^{H}\mathbf{G}\boldsymbol{w}_{i}|^{2}+|u_{k}|^{2}.
\end{equation}
where $u_{k}$ is an auxiliary variable. To solve the problem, we
further introduce an auxiliary variable $m_{k}$, which represents
the positive weight for the MSE. Then $\mathcal{P}_{S}$ is transformed
into an minimum mean-square estimation error (MMSE) problem:
\begin{subequations}
\begin{equation}
\mathcal{P}_{S}^{1}:\min_{\{\boldsymbol{w}_{k},u_{k},m_{k}\}}\sum_{k\in\mathcal{K}}m_{k}E_{k}-\log m_{k}\label{eq:two-stage problem-1-1}
\end{equation}
\begin{equation}
\textrm{s.t.\ }\sum_{k}||\boldsymbol{w}_{k}||_{2}^{2}\leq P_{\textrm{max}},\label{eq:c1-2-1}
\end{equation}
\end{subequations}

In the long-term sub-algorithm, we need to calculate the gradient
of $\mathbf{W}^{J}(\mathbf{X};\mathcal{H})$ w.r.t $\mathbf{X}$,
which is not easy to calculate in closed-form. Therefore, in this
paper, we propose to use a KDL-based approach, to predict dual variables
and power allocation coefficients of the KKT solution of the MMSE
problem $\mathcal{P}_{S}^{1}$. Specifically, the KKT solution of
$\mathcal{P}_{S}^{1}$ is reconstructed as \cite{xu2025joint}
\begin{equation}
\mathbf{W}^{\textrm{KKT}}=\textrm{diag}(\boldsymbol{\varrho})\left(\mathbf{I}+\hat{\mathbf{H}}\textrm{diag}(\boldsymbol{\lambda})\hat{\mathbf{H}^{H}}\right)^{-1}\hat{\mathbf{H}},
\end{equation}
where $\hat{\mathbf{H}}=\mathbf{H}^{H}\mathbf{G}$ is the effective
channel matrix, $\boldsymbol{\varrho}=[\varrho_{1},...,\varrho_{K}]$
is the power allocation coefficients and $\boldsymbol{\lambda}=[\lambda_{1},...,\lambda_{K}]$
is the dual variables. By doing so, we only need to learn $2K$ dual
variables rather than $N\times K\times2$ complex beamforming coefficients.
Moreover, KDL can benefit from both the optimization-guided structure
and the iteration-free data-driven learning, thus enables fast convergence
of the learning algorithm.

We adopt the Transformer in \cite{xu2025joint} to learn the KKT solution,
where the encoder and the decoder using bidirectional self-attention
mechanisms, and the cross-attention mechanism is adopted between the
encoder and the decoder. The input of the KDL-based approach is the
vectorized user location and pinching location (i.e., CSI) sequence,
and the output is $\boldsymbol{\varrho}$ and $\boldsymbol{\lambda}$;
The loss function is the negative system sum rate, which is calculated
based on $\mathbf{W}^{\textrm{KKT}}$ constructed by the output, and
the encoder as well as the decoder will be jointly trained by minimize
the loss function.

Finally, we can simply use BP to obtain the gradient $\partial_{\mathbf{X}}g$
of each short-term CSI sample to update the long-term pinching beamforming. 

\subsection{Summary of the Two-Timescale Design}

We propose a two-timescale joint transmit and pinching beamforming
design for the maximization of sum rate of a PASS-based downlink multi-user
MISO system, which develops a PDD method to decouple the problem,
and a KDL-based learning approach is proposed to obtain a stationary
point of the short-term transmit beamforming and the long-term stochastic
pinching beamforming optimization is solved by a SSCA method. The
overall algorithm is summarized in Algorithm \ref{two-timescale}.
\begin{algorithm}
\caption{\label{two-timescale}The proposed Two-Timescale Algorithm}

\textbf{Initialization:} Initialize $\mathbf{X}^{0}$; $t=0$.

\textbf{Long-term pinching beamforming:}

\phantom{} \phantom{}{\footnotesize 1}. Obtain a mini-batch $\left\{ \mathcal{H}\left(n_{s}\right)\right\} _{n_{s}=\left\{ 1,...,N_{s}\right\} }$;

\phantom{} \phantom{}{\footnotesize 2}.\textbf{For }each sample\textbf{
do }

\phantom{} \phantom{} \phantom{}\textbf{ }\phantom{}\textbf{ Short-term
transmit beamforming:}

\phantom{} \phantom{} \phantom{} \phantom{} Obtain $\mathbf{W}$
and $\partial_{\mathbf{X}}g$ according to Section \ref{subsec:Short-Term-Transmit-Beamforming}.

\phantom{} \phantom{}{\footnotesize 3}. Update $\mathbf{X}^{t+1}$
according to Algorithm \ref{long-term sub}.

\phantom{} \phantom{}{\footnotesize 4}.\textbf{ $t=t+1$.}

\textbf{End}
\end{algorithm}

\section{Simulation Results \label{sec:Simulation-Results}}

In this section, numerical results are presented to illustrate the
effectiveness of the proposed two-timescale joint transmit and pinching
beamforming algorithm. We consider a downlink MIMO system where the
BS serves $K=\{4,6\}$ single-antenna users. The users are randomly
distributed in a range of $S_{x}\times S_{y}=20\times10\,\textrm{m}^{2}$
area. Each RF chain is connected to a waveguide, and each waveguide
has $L=8$ PAs. The maximum transmission power is set as $P_{max}=\{12,16,20,24,28\}$
dBm, and the noise power is $\sigma^{2}=-90$ dBm. The step sizes
$\rho^{t}$ and $\gamma^{t}$ are set to $10/(10+n)^{0.9}$ and $15/(15+n)$,
respectively. The batch size to train the KDL-based Transformer is
100.

\begin{figure}[tp]
\centering
\begin{centering}
\textsf{\includegraphics[scale=0.45]{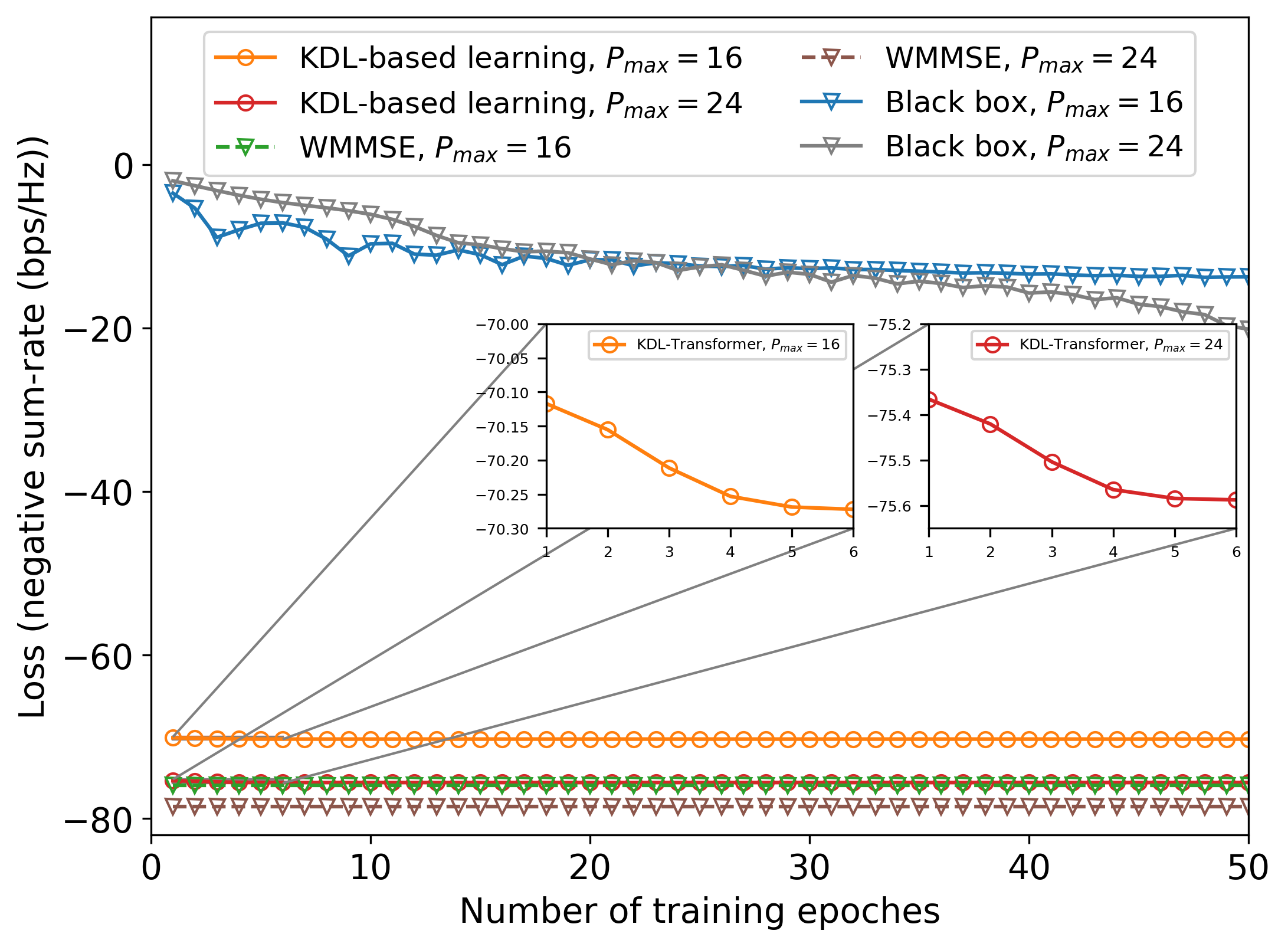}}
\par\end{centering}
\caption{\label{fig:simu1}Comparison of the convergence performance of different
short-term stage algorithms.}
\end{figure}

In Fig. \ref{fig:simu1}, we compare the proposed KDL-based learning
approach in the short-term stage with the conventional WMMSE algorithm
and the black box learning approach, which adopts the end-to-end unsupervised
learning to train a black box for joint transmit and pinching beamforming.
As shown in Fig. \ref{fig:simu1}, the proposed KDL-based learning
approach converges fast in the first several epochs, and converges
to a point close to the near-optimal WMMSE algorithm. The black box
approach performs much worse than the KDL-based learning and the WMMSE.

\begin{figure}[tp]
\centering
\begin{centering}
\textsf{\includegraphics[scale=0.45]{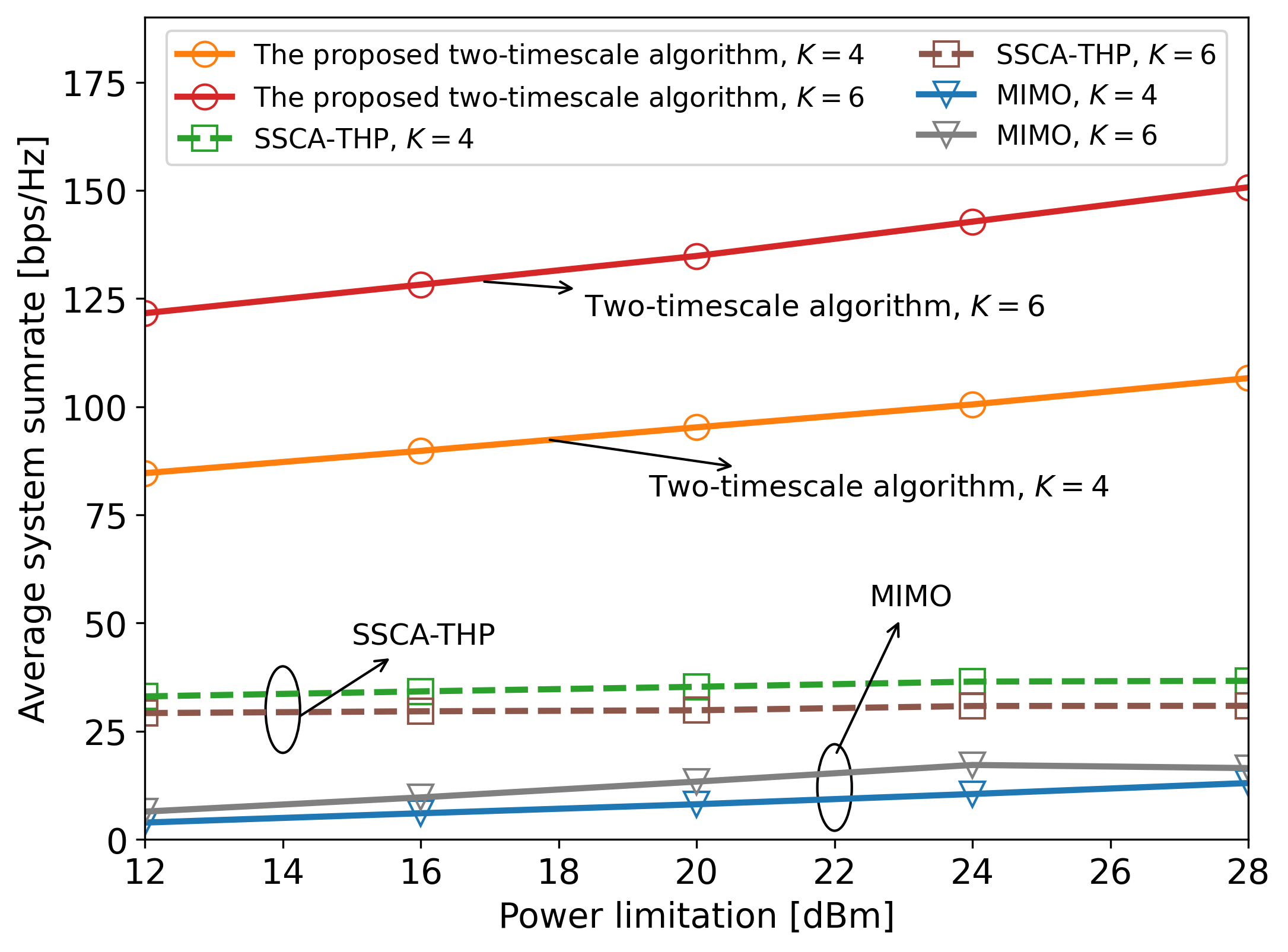}}
\par\end{centering}
\caption{\label{fig:simu2}Performance comparison of different algorithms under
different $P_{max}$.}
\end{figure}

After obtaining the stationary point of the short-term sub-problem
and the corresponding gradient, we can solve the long-term sub-problem.
In Fig. \ref{fig:simu2}, we compare the proposed two-timescale joint
transmit and pinching beamforming with two baselines:
\begin{itemize}
\item \textbf{SSCA-THP}: The SSCA-THP in \cite{liu2019stochastic} is a
single-timescale optimization algorithm which only optimizes the long-term
variable (pinching beamforming in this paper) with the baseband precoder
fixed as the regularized zero-forcing (RZF) precoder.
\item \textbf{MIMO}: The conventional MIMO are equipped with a uniform linear
array comprising $N$ antennas, each connected to a dedicated RF chain.
The corresponding sum rate maximization problem is solved by the WMMSE
algorithm.
\end{itemize}
As shown in Fig. \ref{fig:simu2}, the system sum rate of the proposed
two-timescale joint transmit and pinching beamforming increases as
$P_{max}$ grows while the growth trends of the other two baselines
are not significant. The proposed two-timescale algorithm achieves
a significantly higher performance gain compared to other baselines.
The benefit of PASS-based system is further proved by the performance
gain of the single-timescale optimization based SSCA-THP compared
to the conventional MIMO. 

\section{Conclusion \label{sec:Conclusion}}

This letter proposed a novel two-timescale joint transmit and pinching
beamforming for PASS-based communication systems. The two-timescale
sum rate maximization problem was decoupled by the PDD method into
short-term transmit beamforming sub-problem and long-term pinching
beamforming sub-problem. To tackle the short-term transmit beamforming
and to obtain the gradient, a KDL-based learning approach was proposed,
and the long-term pinching beamforming was solved based on the short-term
stage solution by developing the SSCA. Numerical results have been
provided to verified the efficiency of the proposed two-timescale
algorithm.

\bibliographystyle{IEEEtran}
\bibliography{IEEEabrv,IEEEexample,refer}

\end{document}